\documentclass[usenatbib, A4paper]{mn2e}

\usepackage{amssymb,amsmath}
\usepackage{graphicx}
\usepackage{psfrag}
\usepackage{color}
\title{On the formation of very metal-poor stars: The case of SDSS$\,$J1029151$+$172927}

\author[Klessen, Glover \& Clark]{Ralf S.\ Klessen, Simon C.\ O.\ Glover, Paul C.\ Clark\\
Zentrum f\"{u}r Astronomie der Universit\"{a}t Heidelberg, Institut f\"{u}r Theoretische Astrophysik, Albert-Ueberle-Stra{\ss}e 2, \\69120 Heidelberg, Germany}

\begin{document}
\maketitle

\begin{abstract}
The formation of stars is a key process in the early universe with far reaching consequences for further cosmic evolution. While stars forming from truly primordial gas are thought to be considerably more massive than our Sun, stars in the universe today have typical masses below one solar mass. The physical origin of this transition and the conditions under which it occurs are highly debated. There are two competing models, one based on metal-line cooling as the primary agent and one based on dust cooling. The recent discovery of the extremely metal poor star SDSS$\,$J1029151$+$172927 provides a unique opportunity to distinguish between these two models. Based on simple thermodynamic considerations we argue that SDSS$\,$J1029151$+$172927 was more likely formed as a result of dust continuum cooling rather than cooling by metal lines. We conclude that the masses of extremely metal-poor stars are determined by dust-induced fragmentation.
\end{abstract}

\section{Introduction}
\label{sec:intro}

The earliest generations of stars ended the so-called cosmic dark ages and played a key role in the metal enrichment and reionization of the Universe, thereby shaping the galaxies we see today \citep{Bromm:2004p51274,Glover:2005p514,2005SSRv..116..625C,Bromm:2009p16906}. It is therefore remarkable that there is still very little consensus on the physical processes that govern stellar birth in the early universe. In particular, there is still a debate over how and when the star formation process switches from the high-mass mode  that is thought to dominate in the very early universe \citep{Abel:2002p16262,Bromm:2002p51374,Yoshida:2006p23342,OShea:2007p22165} to the low-mass mode that presides today \cite[e.g.,][]{Kroupa:2002p20470, Chabrier:2003p27882,Bastian:2010p49383}.

The relatively new field of ``stellar archeology" \citep{Beers:2005p549} attempts to address this issue by compiling observational databases of stars with extremely low metallicities in the hope that we can start to constrain the various theoretical predictions. The recent discovery of the extremely metal poor star SDSS$\,$J1029151$+$172927 by \citet{Caffau:2011p51105} provides a unique opportunity for this task. We recap the basic properties of this star in Section \ref{sec:star}. In Section~\ref{sec:SF}, we review the two main competing models proposed to lie behind the transition from the primordial to the present-day 
initial mass function (IMF). These are fragmentation due to metal-line cooling and due to dust continuum cooling. We present simple thermodynamic arguments in Section \ref{sec:comp} and argue that SDSS$\,$J1029151$+$172927 was more likely formed as a result of dust continuum cooling, rather than cooling by metal lines. Finally, we conclude in Section \ref{sec:conclusion}. 

\section{Properties of SDSS$\,$J1029151$+$172927}
\label{sec:star}

SDSS$\,$J1029151$+$172927 is a Galactic halo star at the position $\alpha = 10^h 29^m 15.15^s$ and $\delta = +17^\circ 29' 28''$ (epoch 2000) with a $g$-band magnitude of 16.92$^m$ and an estimated mass of about $0.65\,$M$_{\odot}$ (Caffau et al., in preparation). It was selected from the Sloan Digital Sky Survey and spectral information was obtained with the X-Shooter and UVES instruments at the Very Large Telescope of the European Southern Observatory in Chile.\footnote{SDSS: {\tt www.sdss.org/},  X-Shooter: {\tt www.eso.org/\-sci/\-facilities/\-paranal/\-instruments/\-xshooter/}, UVES: {\tt www.eso.org/\-sci/\-facilities/\-paranal/\-instruments/\-uves/}. }

The star is remarkable in the sense that it is the first star to be discovered with elemental abundances
in the range  $\lesssim10^{-5}$ to $10^{-4}$  of the solar value for all of the elements measured in its spectrum.
This sets it apart from other extremely metal-poor stars with [Fe/H]$\sim -5$ which typically have enhanced CNO abundances, such that [C/H], [N/H], and [O/H] are all at or above values 
of roughly $-3.5$ \citep[e.g.][]{2004ApJ...603..708C,2004ApJ...612L..61B,2007ApJ...670..774N,2008ApJ...684..588F}.\footnote{Note that \citet{Caffau:2011p51105} report no direct detection of oxygen, because there was no suitable line in the spectral range considered. They argue, however, that the spectral fit based on one- and three-dimensional model atmospheres is most consistent when assuming a `standard' value of [O/Fe] of $+0.6$, implying that $[{\rm O}/{\rm H}] \sim -4.4$.}

\section{High-Redshift Star Formation}
\label{sec:SF}

\subsection{Metal-free Stars}
\label{subsec:first}
The first generation of stars, the so-called population III (or Pop III) build up from truly metal-free primordial gas. They have long been thought to live short, solitary lives, with only one extremely massive star with typically $\sim 100\,$M$_\odot$ or more forming in each dark matter minihalo \citep{Abel:2002p16262,Bromm:2002p51374,Yoshida:2006p23342,OShea:2007p22165}. However, in more recent simulations \citet{Turk:2009p20571} and \citet{Stacy:2010p51133} have reported fragmentation of primordial gas into a wide binary system for some of their cases considered. Widespread fragmentation into multiple stellar systems and clusters has also been found by \citet{Clark:2008p5179, Clark:2011p44986}, \citet{Smith:2011p47178} and \citet{Greif:2011p51156}. Although these calculations suggest an IMF for primordial stars that reaches down to masses perhaps as low as $0.5\,$M$_{\odot}$, the primary mode of star formation is expected to lead to higher-mass objects with typically several tens of solar masses, as the emerging mass function appears to be flat \citep{Clark2011a,Greif:2011p51156}. This agrees with the analysis of abundance patterns of extremely metal-poor stars in the Galactic halo \cite[e.g.,][]{Beers:2005p549}, which are consistent with the enrichment from core collapse supernovae from stars in the mass range $20 - 40\,$M$_{\odot}$ rather than from pair-instability supernovae of progenitors of $\sim 200\,$M$_{\odot}$ \citep{Tumlinson:2004p32682, Tumlinson:2007p9256,Tominaga:2007p51761, Izutani:2009p51634, Heger:2010p51925, Joggerst:2010p27822}. In addition, the metal enrichment of the IGM occurs relatively rapidly following the onset of Pop III star formation \citep[see e.g.][]{greif10,maio11,wise11}, significantly limiting the total number of Pop III stars that are able to form.

\subsection{Metal-Enriched Stars}
\label{subsec:second}
The second generation of stars, often termed Pop II.5, forming from material that has been enriched from the debris of the first stars is the topic of this discussion here. There are two main models for the physical processes that determine the fragmentation behavior of low-metallicity gas and consequently set the stellar mass spectrum. One model is based on cooling by atomic lines and the other one by dust. Both predict  that increasing metallicity leads to a transition from a primordial mode of star formation, producing predominantly higher-mass stars (even though the characteristic mass of this mode is subject to debate as discussed above), to the well-known present-day mode of star formation, where the IMF peaks at around $0.2\,$M$_{\odot}$ \cite[e.g.,][]{Kroupa:2002p20470, Chabrier:2003p27882,Bastian:2010p49383}. However, there is disagreement on the physical origin of this transition and on the critical metallicity $Z_{\rm crit}$ where it occurs. The discovery of the very primitive star SDSS$\,$J1029151$+$172927 \citep{Caffau:2011p51105} opens up the unique opportunity, to distinguish between these two competing models and to determine which of these is more consistent with the available observational data. 

\vspace{0.2cm}\noindent{\bf Metal-Line Cooling} ---
The first model focuses on the cooling by atomic fine-structure lines from alpha elements such as carbon or oxygen at densities around $10^4\,$cm$^{-3}$. This leads to $Z_{\rm crit} \sim 10^{-3}\,Z_{\odot}$, which can be estimated analytically by calculating the metallicity required to produce a cooling rate equal to the rate of adiabatic compression heating for given halo properties \citep{Bromm:2003p51252, Santoro:2006p52095}. This lead \citet{Frebel:2007p52034} to combine the C and O abundance measures and introduce a transition discriminant $D_{\rm trans} = \log_{10} \left( 10^{\rm [C/H]} + 0.3 \times 10^{\rm [O/H]}\right)$. They propose that low-mass stars can only form for $D_{\rm trans} \gtrsim -3.5$. This proposition is challenged by the discovery of  SDSS$\,$J1029151$+$172927, for which $D_{\rm trans} = -4.2$, if we use the upper limit on [C/H] derived by \citet{Caffau:2011p51105} and also follow them in assuming that $[{\rm O}/{\rm Fe}] = +0.6$ for this star. We note that larger $D_{\rm trans}$ values are possible if the oxygen abundance is larger than assumed. However, $D_{\rm trans} = -3.5$  would require $[{\rm O}/{\rm Fe}] \geq +1.9$, which seems highly unlikely given that C, N and other $\alpha$ elements do not show signs of significant enhancement.

\vspace{0.2cm}\noindent{\bf Dust-Induced Fragmentation} --- 
Another line of reasoning considers dust cooling as the primary agent for fragmentation \citep{Omukai:2005p20777,Omukai:2010p42237, Schneider:2002p21947, Schneider:2006p21948, Schneider:2011p52132}. 
In this model, $Z_{\rm crit}$ is between $10^{-5}$ to $10^{-6}\,Z_{\odot}$ where much of the uncertainty in the predicted value comes from our poor knowledge of the dust composition and the degree of gas-phase depletion \cite[e.g.,][]{Schneider:2011p52132}.  Detailed numerical simulations based on a tabulated equation of state (EOS) by \citet{Tsuribe:2006p23371,Tsuribe:2008p23366}  and  \citet{Clark:2008p5179} demonstrated that indeed fragmentation occurs at densities around and above $n\approx 10^{12}\,$cm$^{-3}$ and at metallicities $10^{-5} - 10^{-6}\,Z_{\odot}$, leading to an IMF with a peak below $1\,$M$_{\odot}$ consistent with the present-day IMF.  This approach has been refined by \citet{Dopcke:2011p41016} who performed similar calculations with time-dependent chemistry and found equivalent results.

\section{Metal-Line vs. Dust Cooling: the Case of SDSS$\,$J1029151$+$172927}
\label{sec:comp}

\subsection{Basic Theoretical Considerations}
\label{subsec:basics}

Our current understanding of stellar birth at the present day is based on the intricate interplay between the self-gravity of the star-forming gas and various opposing physical agents, such as pressure gradients, magnetic fields, or turbulence \cite[see, e.g., the reviews by][]{MacLow:2004p2713,Mckee:2007p34,BallesterosParedes:2007p6136}. Despite these complexities, the characteristic mass of the stars that form appears to be tightly controlled by the thermodynamic state of the gas \cite[e.g.,][]{2000ApJS..128..287K,2002ApJ...576..870P,Larson:2005p657, Jappsen:2005p45068, 2006MNRAS.368.1787C, 2009ApJ...702.1428H}. That means that the thermodynamic properties of the star-forming gas, i.e.\ the balance between heating and cooling processes, are the key to understanding star formation in the primordial universe. 

In a simple numerical experiment \citet{Li:2003p6067} showed that turbulent gas with an
equation of state $P \propto n^{\gamma}$ (where $P$ is the pressure and $n$ is the number density of the gas) could fragment efficiently and build up a cluster of low mass stars provided that $\gamma < 1$.
Conversely, star formation in gas with $\gamma >1$ is biased towards forming very few and very massive objects. \citet{Larson:1985p52413,Larson:2005p657} proposed that the transition from a cooling-dominated to a heating-dominated regime (i.e.\ the transition from $\gamma < 1$, which implies a decreasing temperature with increasing density,  to $\gamma > 1$, which implies a rising temperature with rising density) marks the point at which the gas fragments and therefore is responsible for determining a characteristic mass scale for fragmentation. Based on this proposal, \citet{Jappsen:2005p45068} performed a suite of numerical simulations using piece-wise polytropic equations of state and demonstrated that Larson's proposal appears to be correct and that the 
transition from $\gamma < 1$ to $\gamma > 1$ does indeed define a characteristic mass scale for fragmentation \citep[see also][]{2006MNRAS.368.1296B}.

Physically, this behaviour can be understood as a consequence of fragmentation in turbulent, self-gravitating gas occurring primarily in filaments. Filaments with $\gamma \leq 1$ are unstable to fragmentation \citep{Larson:1985p52413,Larson:2005p657}, while those with $\gamma > 1$ are stable and do not fragment. \citet{Kawachi:1998p52472} showed when studying the similarity solution for the evolution of self-gravitating gaseous filaments that the collapse becomes slower and slower as $\gamma$ approaches unity from below and eventually comes to a halt at $\gamma = 1$. They suggested that the slow-down of the collapse as $\gamma$ approaches unity will in reality cause a filament to fragment into clumps, because the timescale for fragmentation then becomes shorter than the timescale for contraction toward the axis of the filament. Simulations of star formation in present-day molecular clouds indicate that most stars do indeed form within filaments \citep[e.g.][]{2011MNRAS.411.1354S}, and this idea also has strong observational support \citep[see e.g.][]{2011A&A...529L...6A}. As numerical models of the assembly of the first galaxies show that the gas develops a filamentary morphology \citep[e.g.][]{2008ApJ...685...40W, 2008MNRAS.387.1021G, Bromm:2009p16906}, it seems reasonable to assume that filament fragmentation will dominate in the earliest galaxies just as it does in local star-forming regions.

\begin{figure}
\begin{center}
\includegraphics[width=0.999\columnwidth]{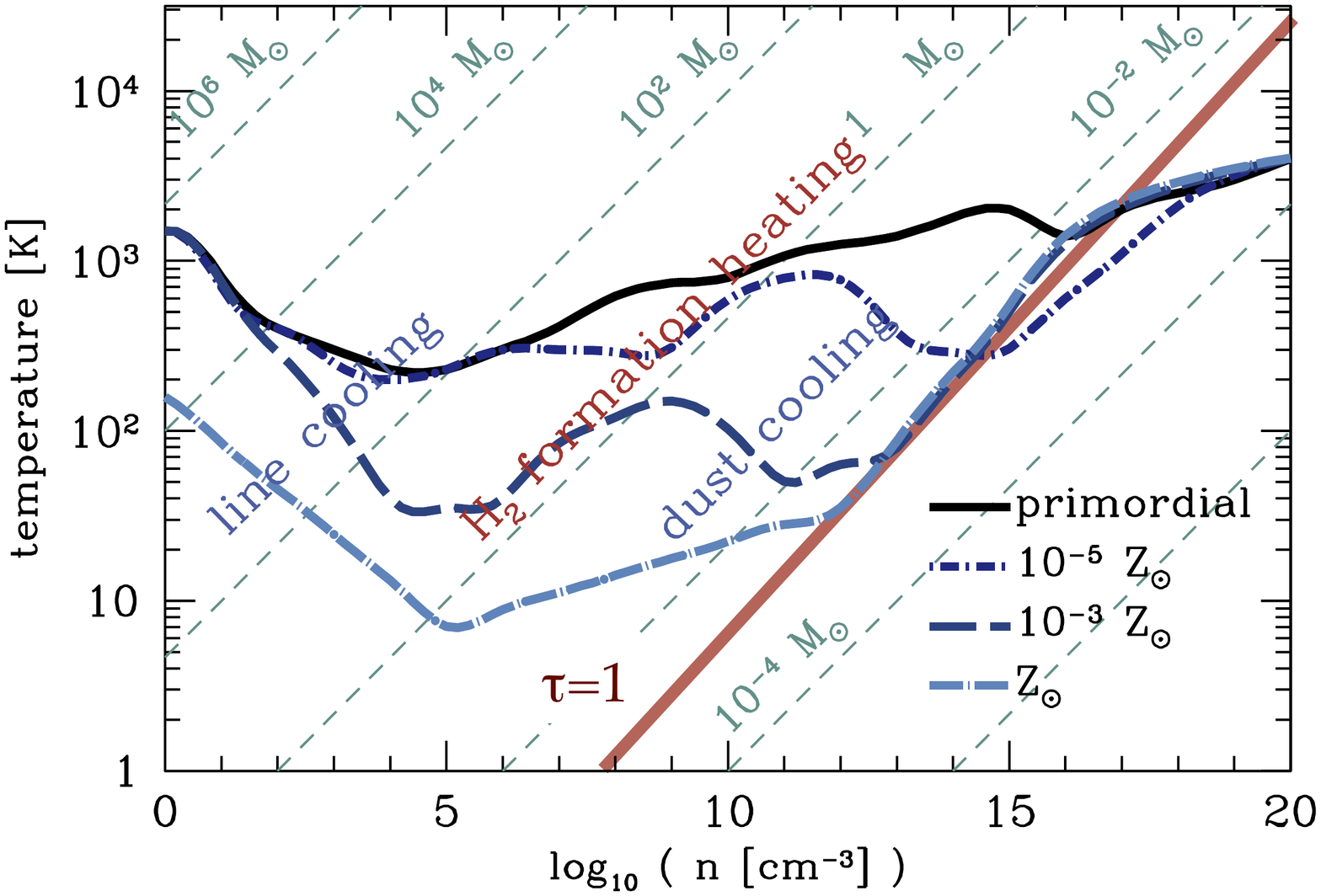}
\caption{A schematic diagram showing the relation between the expected equilibrium temperature $T$ and the density $n$ for different levels of metal enrichment, $Z/Z_{\odot} = 0, 10^{-5}, 10^{-3}$, and $1$. The `dip' in the $T-n$ relation at low densities is due to metal-line cooling followed by H$_2$ formation heating, while the `dip' at high densities and subsolar metallicities  is due to dust cooling followed by the transition to the optically thick regime ($\tau > 1$, solid red line) for continuum radiation. Thin dashed lines indicate the lines of constant Jeans mass. The $T-n$ relation shown here is representative of the results from detailed studies. The solar metallicity line for $n \lesssim 10^{12} \rm cm^{-3}$ is an approximation proposed by \citet{Larson:2005p657}, combined with the expected shift to $\gamma \approx 1.4$ at a density of $10^{12} \rm cm^{-3}$, where  the gas become optically thick (e.g.\ \citealt*{Masunagaetal1998}). Above $n \approx 10^{15}\,$cm$^{-3}$ the temperature is high enough to dissociate H$_{2}$ leading to $\gamma \sim 1.1$. The other three curves for the lower metallicity and primordial gas reflect the one-zone calculations by \citet{Omukai:2005p20777,Omukai:2010p42237}; see also \citet{Schneider:2002p21947, Schneider:2006p21948, Schneider:2011p52132}, or \citet{Dopcke:2011p41016} for a three-dimensional calculation. 
}
\label{fig:EOS}
\end{center}
\end{figure}

\subsection{Application to Metal-Poor Gas}
\label{subsec:application}
The theoretical picture outlined above suggests that in order to make predictions for 
the characteristic mass of stars formed in low metallicity gas, we need to establish the conditions in
which the effective equation of state changes from $\gamma < 1$ to $\gamma > 1$. The locations where such `dips' in the temperature-density ($T-n$) relation occur are illustrated in Figure \ref{fig:EOS}. Note that for an ideal gas with a barotropic EOS (where $P \propto n^\gamma$) we obtain the relation $\gamma = d \ln T / d \ln n + 1$ and can simply read off the effective index $\gamma$ from the slope of the $T-n$ relation. Two of sets of features are particularly important for the thermodynamic analysis here. The first sequence of `dips' is caused by metal-line cooling. This can be highly effective at low densities, but becomes much less effective for densities $n > n_{\rm crit}$, the critical density at which the fine structure level populations reach their local thermodynamic equilibrium values, leading to an increase in the temperature. The growing effectiveness of H$_{2}$ formation heating as we move to higher densities also drives $T$ to higher values. As shown by \citet{Omukai:2005p20777,Omukai:2010p42237}, this feature covers the metallicity range $0 \,\lesssim \,$[M/H]$\,\lesssim\, -4$ (where ${\rm M}$ denotes the total metal abundance). We see from Figure \ref{fig:EOS} that minima in the $T-n$ curve induced by metal-line cooling correspond to Jeans masses significantly above $10\,$M$_\odot$. Indeed, \citet{Bromm:2001p51382} and \citet{Jappsen:2009p837}, who directly simulate fragmentation in this regime, report minimum fragment masses in excess of $10\,$M$_\odot$, with the majority of fragments having masses of order  $100\,$M$_\odot$. These calculations demonstrate that the line-cooling regime is unable to directly produce stars with masses around 1 $\rm M_{\odot}$ for metallicities of $[{\rm M}/{\rm H}]\sim -3$ and below \cite[see also][]{Smith2009}. This is the typical mass range of the extremely metal poor stars observed so far \citep{Beers:2005p549}. We conclude that metal-line cooling cannot explain the physical characteristics of these highly interesting objects. 

The second sequence of `dips' in Figure~\ref{fig:EOS} is caused by the coupling between gas and dust and covers a range of metallicities from the solar value to about $Z \approx 10^{-5}\,Z_\odot$ or possibly lower if different dust models are considered. It occurs at densities just below the critical value for the gas to become optically thick for continuum radiation as indicated by the red line. It is important to note that because this sequence of 'dips' lies at much higher densities, the corresponding characteristic mass for fragmentation falls significantly below $1\,$M$_\odot$. Based on this simple thermodynamic argument, \citet{Omukai:2005p20777,Omukai:2010p42237}, \citet{Schneider:2006p21948,Schneider:2011p52132}, and others concluded that the observed extremely metal-poor stars in the Galactic halo must be caused by dust-induced fragmentation rather than by metal-line cooling. This proposition was confirmed in numerical simulations by \citet{Tsuribe:2006p23371,Tsuribe:2008p23366},  \citet{Clark:2008p5179}, and \citet{Dopcke:2011p41016}. These calculations show that dust-cooling at high densities frequently leads to objects in the sub-solar mass range down to metallicities of [M/H]$\, \sim -5$. This is exactly the physical regime represented by the recently discovered halo star SDSS$\,$J1029151$+$172927. 

A potential objection against this model is that it is possible that dust production in very metal poor gas is very inefficient \citep[see e.g.][]{sss10}, although this is still quite uncertain. Nevertheless, if
this were the case, then the high-density `dip' would disappear, and the dynamical behavior of the gas would be very similar to the zero-metallicity case, because metals in the gas phase contribute little to the overall cooling rate for [M/H]$\,\lesssim\, -4$ and densities above $n \sim 10^{10}\,$cm$^{-3}$ \citep{Omukai:2005p20777,Omukai:2010p42237}. As discussed in Section \ref{subsec:first}, even zero-metallicity gas is able to form low-mass stars via disk fragmentation at high densities  \citep{Clark:2011p44986, Greif:2011p51156}. Consequently, SDSS$\,$J1029151$+$172927 could in principle have formed without dust cooling. However, this formation channel is thought to be rare and the additional cooling provided by dust is needed to build up sufficiently large numbers of low-mass stars for the mass spectrum to be consistent with the present-day IMF with its steep negative slope from small to large masses (Dopcke et al., in preparation). Therefore it would provide additional support for the importance of dust fragmentation for star formation in the early universe if stars like SDSS$\,$J1029151$+$172927 were found to be common in our Milky Way. The current attempts to survey larger numbers of very metal poor stars will be able to address this issue. 

\section{Conclusions}
\label{sec:conclusion}

In this Letter we have argued that the physical characteristics (mass and metallicity) of the extremely metal-poor halo star SDSS$\,$J1029151$+$172927 are consistent with it having formed via dust-induced fragmentation in the high redshift universe, but are inconsistent with the proposal of fragmentation induced by metal-line cooling. 

The fact that \citet{Caffau:2011p51105} find one extremely metal poor star with $[{\rm M}/{\rm H}] \gtrsim -5$ out of a sample of only six candidate objects suggests that this type of star may not be as uncommon as previously thought. With about 2,900 additional candidate stars in the 7$^{\rm th}$ SDSS data release, the chances of finding more such metal-poor stars seems to be quite high. 

A number of previous studies have attempted to compute the number of low metallicity, low mass stars that one would expect to find in the Galactic halo at the present day \citep[see e.g.][]{2006ApJ...653..285S,2007MNRAS.381..647S, 2010MNRAS.401L...5S}. To derive a very crude estimate, we note that of order $10^4$ minihalos with metallicities below $10^{-5}\,Z_{\odot}$ contributed to the early assembly history of the Milky Way \citep{Diemand:2005p52477,Gao:2010p52208}, and that the calculations by \citet{Clark:2008p5179} and \citet{Dopcke:2011p41016} typically produce ten or more low-mass protostars with masses below $0.8\,$M$_\odot$ (the limit for stars to survive until the present day; see e.g.\ \citealt{2004ApJ...609.1035P}). Although these calculations have a number of serious limitations, if we adopt a value of ten low mass stars per halo as a first approximation, then we estimate that there are of order of $10^5$ primitive stars in our Galaxy. It is interesting to note that the value that we obtain from this simple estimate agrees to within a factor of ten with the values predicted by more complex models \citep[e.g.][]{2007MNRAS.381..647S}.

Finally, we note that current models of hierarchical galaxy formation predict that the first and most metal-poor halos to merge will become part of the bulge component of the resulting spiral galaxy \cite[e.g.,][]{Tumlinson:2010p21636},  implying that this may be the most promising location to search for extremely metal-poor and metal-free stars \cite[e.g.,][]{White:2000p52248}. In this context it is interesting to note that the high eccentricity of SDSS$\,$J1029151$+$172927 is consistent with it being ejected from the bulge of the Milky Way (Caffau et al., in preparation).

\section*{Acknowledgements}
{We thank Elisabetta Caffau and Hans-G\"{u}nter Ludwig for providing us with with a mass estimate and kinematic information about SDSS$\,$J1029151$+$172927 prior to publication. We also thank Volker Bromm, Gustavo Dopcke, Naoki Yoshida, Kazuyuki Omukai,  Raffaella Schneider, and Rowan Smith for many stimulating discussions. We are grateful to the referee for helpful comments. The authors acknowledge financial support from the {\em Landesstiftung Baden-W\"{u}rttemberg} via their program {\em Internationale Spitzenforschung II} (grant P-LS-SPII/18), from the German {\em Bundesministerium f\"{u}r Bildung und Forschung} via the ASTRONET project STAR FORMAT (grant 05A09VHA), from the {\em Deutsche Forschungsgemeinschaft} (DFG) under grants KL1358/11-1 and KL1358/14-1, and via the SFB 881 ÒThe Milky Way GalaxyÓ.}


\end{document}